\documentclass[conference]{IEEEtran}

\usepackage{cite}
\ifCLASSINFOpdf
   \usepackage[pdftex]{graphicx}
\else
\fi
\usepackage{amsmath,amssymb,amsthm}
\usepackage{mathrsfs}
\usepackage{amsbsy}
\usepackage{steinmetz}
\usepackage{nicefrac}
\usepackage{soul}
\usepackage{multirow}

\usepackage{tikz}
\usepackage{pgfplots}
\usepgfplotslibrary{units}

\usepackage{array}
\ifCLASSOPTIONcompsoc
  \usepackage[caption=false,font=normalsize,labelfont=sf,textfont=sf]{subfig}
\else
  \usepackage[caption=false,font=footnotesize]{subfig}
\fi

\IEEEoverridecommandlockouts

\newcommand{\fixme}[2]{\ifx&#2&{\color{red}#1}\else{\color{red}FIXME\{}#1{\color{red}\}}\footnote{{\color{red}#2}}\PackageWarning{Fixme}{#1: #2}\fi}
\usepackage{booktabs}
\usepackage{color}

\definecolor{Set1-7-1}{RGB}{228,26,28}
\definecolor{Set1-7-2}{RGB}{55,126,184}
\definecolor{Set1-7-3}{RGB}{77,175,74}
\definecolor{Set1-7-4}{RGB}{152,78,163}
\definecolor{Set1-7-5}{RGB}{255,127,0}
\definecolor{Set1-7-6}{RGB}{166,86,40}
\definecolor{Set1-7-7}{RGB}{0,0,0}

\newcommand{\figurewidth}{0.95}	
\newcommand{\figureheight}{0.76}	

\begin{document}

\bstctlcite{IEEEexample:BSTcontrol}

\title{Coded LoRa Frame Error Rate Analysis}
\author{\IEEEauthorblockN{Orion Afisiadis\IEEEauthorrefmark{1}, Andreas Burg\IEEEauthorrefmark{1}, and Alexios Balatsoukas-Stimming\IEEEauthorrefmark{2}}\\
\IEEEauthorblockA{\IEEEauthorrefmark{1}Telecommunication Circuits Laboratory, \'{E}cole polytechnique f\'{e}d\'{e}rale de Lausanne, Switzerland\\
\IEEEauthorrefmark{2}Department of Electrical Engineering, Eindhoven University of Technology, Netherlands}
}

\maketitle

\begin{abstract}
In this work, we study the coded frame error rate (FER) of LoRa under additive white Gaussian noise (AWGN) and under carrier frequency offset (CFO). To this end, we use existing approximations for the bit error rate (BER) of the LoRa modulation under AWGN and we present a FER analysis that includes the channel coding, interleaving, and Gray mapping of the LoRa physical layer. We also derive the LoRa BER under carrier frequency offset and we present a corresponding FER analysis. We compare the derived frame error rate expressions to Monte Carlo simulations to verify their accuracy.
\end{abstract}

\IEEEpeerreviewmaketitle

\section{Introduction} \label{sec:intro}

LoRaWAN is a very popular communications protocol for the Internet of things (IoT). Its physical layer (PHY), which is called LoRa, is based on a proprietary spread spectrum modulation scheme that uses chirp modulation as its basis~\cite{SX127x}. Along with the chirp modulation, the LoRa PHY chain includes whitening, channel coding, interleaving and Gray mapping~\cite{Ghanaatian2019}. LoRa is able to work in a wide range of operational signal-to-noise ratios (SNRs), due to the support of multiple spreading factors (SF) and code rates. Some of the details of the LoRa PHY have been revealed in patents~\cite{Seller2016}, but also through several reverse-engineering efforts~\cite{Knight2016,Robyns2018}. 

There are several works in the literature that derive approximate formulas for the bit error rate (BER) and the symbol error rate (SER) of the LoRa modulation under both additive white Gaussian noise (AWGN)~\cite{Reynders2016b,Elshabrawy2018} and under Rayleigh fading channels~\cite{Elshabrawy2018}. Moreover, the work of~\cite{Elshabrawy2018b} is the first to give a low-complexity SER approximation under AWGN and interference from a chip-aligned LoRa interferer with the same spreading factor (same-SF interference) as the user of interest. This study was extended to the more realistic non-chip-aligned and non-phase-aligned interference scenario in~\cite{Afisiadis2019,Afisiadis2019b}. The work in~\cite{Afisiadis2019b} also contains an approximation of the FER under AWGN and same-SF interference, which can be particularly useful for LoRa network simulators (e.g., \cite{Markkula2019,Reynders2018,Pop2017,Abeele2017,Bor2016}) but also for theoretical analyses (e.g., \cite{Sorensen2019,Amichi2019}) instead of hard reception thresholds, which have been experimentally shown to be too simplistic~\cite{Fernandes2019}. However, the FER expression in~\cite{Afisiadis2019b} does not take into account the channel coding, interleaving, and Gray mapping present in a LoRa transceiver physical layer chain, which have a great impact on the error rate. The FER expression in~\cite{Afisiadis2019b} also does not take into account any potential PHY impairments, which are very important in the context of low-power radios.

\subsubsection*{Contributions}
In this work, we analyze the channel coding and interleaving mechanisms of the LoRa PHY in order to first derive the codeword error rate (CWER) for LoRa. We then use the CWER to derive two low-complexity approximations for the coded FER of a LoRa system under AWGN. Moreover, we derive an approximation for the coded FER of a LoRa system under AWGN and residual carrier frequency offset (CFO), which is an important impairment that can significantly affect the performance of LoRa. Finally, we corroborate the accuracy of our approximations through Monte Carlo simulations.

\subsubsection*{Notation}
We denote the probability density function (PDF) and the cumulative density function (CDF) of the Rayleigh and Rice distributions by $f_{\text{Ra}}(y; \sigma)$, $f_{\text{Ri}}(y; v, \sigma)$ and $F_{\text{Ra}}(y;\sigma)$, $F_{\text{Ri}}(y;v, \sigma)$, respectively, where $\sigma$ and $v$ are the \emph{scale} and \emph{location} parameters. Moreover, bold lowercase letters (e.g., $\mathbf{a}$) denote vectors and bold uppercase letters (e.g., $\mathbf{A}$) denote matrices. Bold calligraphic letters (e.g., $\pmb{\mathcal{A}}$) denote a vector containing the frequency-domain representation of $\mathbf{a}$, i.e., $\pmb{\mathcal{A}} = \text{DFT}(\mathbf{a})$, while regular calligraphic letters denote sets (e.g., $\mathcal{A}$). The $i$-th element of a vector $\mathbf{v}$ is denoted by $\mathbf{v}_i$.

\section{LoRa PHY System Model} \label{sec:system_model}

In this section, we provide some background on the LoRa modulation and demodulation procedures, as well as on the structure of a packet used in a LoRa data transmission.

\subsection{LoRa Modulation and Demodulation}
LoRa is a spread-spectrum frequency modulation that uses a bandwidth $B$ and $N = 2^\text{SF}$ chips per symbol, where $\text{SF}$ is called the \emph{spreading factor} and $\text{SF} \in \{7, \dots, 12\}$. Each LoRa symbol carries $\text{SF}$ bits of information and Gray mapping is used to convert the bits to symbols. A baseband symbol $s \in \mathcal{S}$, where $ \mathcal{S} = \left\{0,\hdots,N{-}1\right\} $, begins at frequency $(\frac{s B}{N} - \frac{B}{2})$, and its frequency increases by $\frac{B}{N}$ for each chip until the Nyquist frequency $\frac{B}{2}$ is reached, at which point a frequency fold to $-\frac{B}{2}$ occurs. The general discrete-time baseband equivalent description of a LoRa symbol $s$, for the common case where the sampling frequency $f_s$ is equal to $B$, is~\cite{Afisiadis2019}
\begin{align} \label{eq:LoRa_symbol}
x[n] & = e^{j2\pi \left(\frac{n^2}{2N}  + \left(\frac{s}{N} - \frac{1}{2}\right)n \right)}, \;\; n \in \mathcal{S}.
\end{align}
When transmission takes place over an AWGN channel, the received LoRa symbol is given by
\begin{align}
  y[n] & = x[n] + z[n], \;\; n \in \mathcal{S}, \label{eq:lora_rx}
\end{align}
where $z[n] \sim \mathcal{CN}(0,\sigma^2)$ is complex AWGN with variance $\sigma^{2} = \frac{N_0}{N}$ and singled-sided noise power spectral density $N_0$. The SNR is defined as $\text{SNR} = \frac{1}{N_0}$.

\begin{figure}[t]
	\centering
	\includegraphics[width=0.5\textwidth]{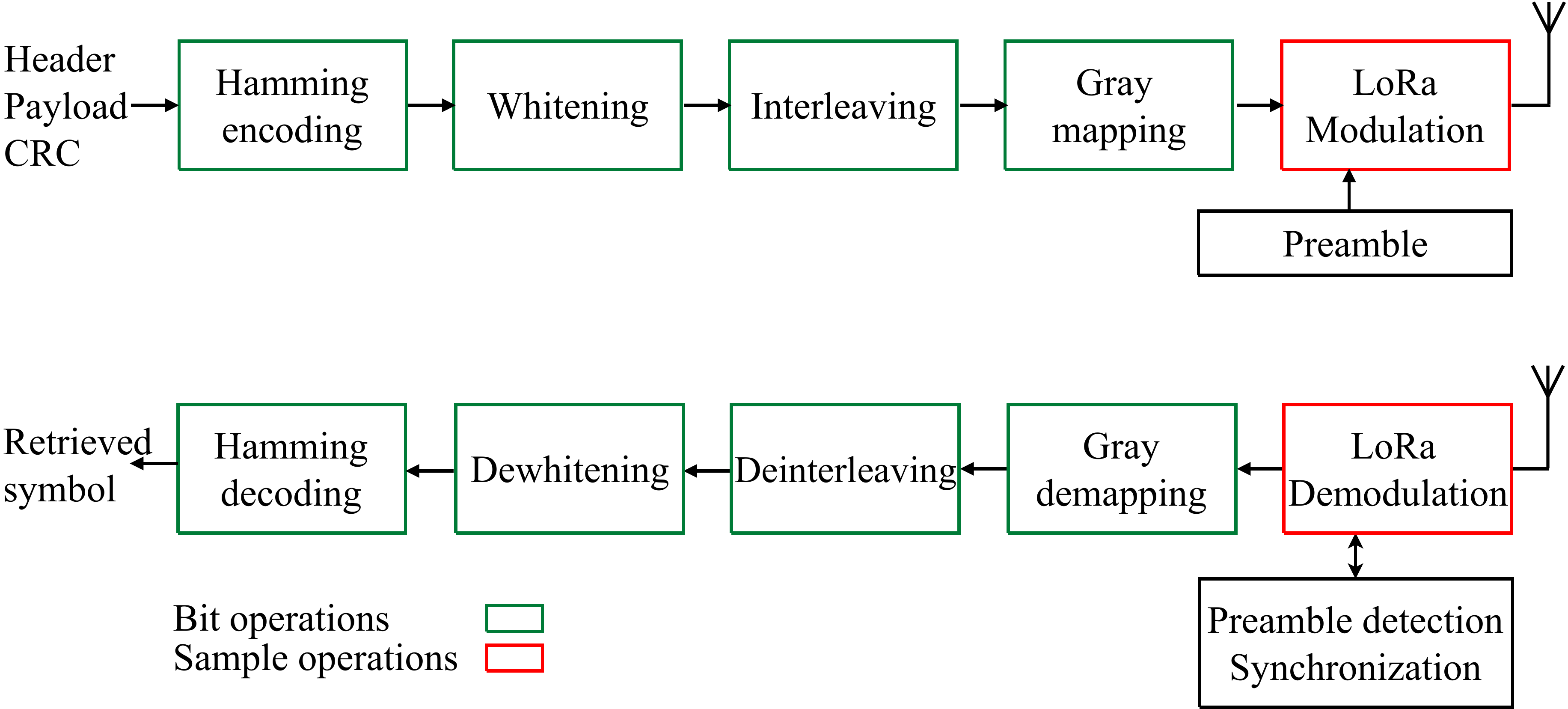}
	\caption{Illustration of LoRa PHY Tx and Rx chains.}
	\label{fig:blockDiagram}
	\vspace{-0.25cm}
\end{figure}

The receiver first \emph{dechirps} the received signal through an element-wise multiplication of $y[n]$ with the complex conjugate of an \emph{upchirp} reference signal $x_{\text{ref}}[n]$ (i.e., the LoRa symbol for $s=0$). As a next step, the receiver computes the discrete Fourier transform (DFT) of the dechirped signal and obtains $\pmb{\mathcal{Y}} = \text{DFT}\left(\mathbf{y} \odot \mathbf{x}_{\text{ref}}^{*}\right)$, where $\odot$ denotes the Hadamard product, $\mathbf{y} = \begin{bmatrix} y[0] & \hdots & y[N-1] \end{bmatrix}$, and $\mathbf{x}^*_{\text{ref}} = \begin{bmatrix} x^*_{\text{ref}}[0] & \hdots & x^*_{\text{ref}}[N-1] \end{bmatrix}$. An estimate $\hat{s}$ of the transmitted LoRa symbol is finally obtained as
\begin{align}
  \hat{s} = \arg\max_{k\in \mathcal{S} } \left( |\pmb{\mathcal{Y}}_k| \right). \label{eq:retrieved_symbol_dft}
\end{align}
The $\text{SF}$-bit label of $\hat{s}$, which is defined in the LoRa standard, corresponds to an estimate of the $\text{SF}$ transmitted bits.

\subsection{LoRa Transceiver Chain}\label{sec:block_diagram}
As shown in Fig.~\ref{fig:blockDiagram} the transmitter chain of LoRa includes a Hamming encoder, a whitening block, an interleaver, and a Gray-mapping block prior to the chirp modulation~\cite{Robyns2018}. The receiver contains a Gray demapper, a deinterleaver, a dewhitening block, and a Hamming decoder.

LoRa uses $(k,n)$ Hamming codes with $ k = 4 $ and $n \in \{5,6,7,8\}$, where $k$ denotes the data word length and $n$ denotes the codeword length. The $(4,5)$ and $(4,6)$ Hamming codes can only detect errors, while the $(4,7)$ and $(4,8)$ Hamming codes can correct one bit-error per codeword. Since the $(4,5)$ and $(4,6)$ Hamming codes are effectively uncoded with respect to the error rate, we focus our analysis on the $(4,7)$ and $(4,8)$ Hamming codes. The bits included in a block of SF codewords are interleaved using a diagonal interleaver. The combination of the Hamming code and the interleaving has a great effect on the error rate and is discussed in more detail in Section~\ref{sec:FER}. 
The whitening, on the other hand, has no effect the error rate and can thus be ignored. In the presence of only AWGN, all symbol errors are equally likely, meaning that the effect of Gray mapping can also be ignored when deriving the FER. However, in the presence of AWGN and CFO the pair-wise error probability for symbol $s$ is non-uniform and adjacent-symbol errors are significantly more likely~\cite{Ghanaatian2019}, in which case the Gray mapping plays a very important role and can thus not be ignored.

\subsection{LoRa Packet Structure}
The structure of a LoRa packet as explained in~\cite{Robyns2018} is illustrated in Fig.~\ref{fig:blockDiagram}. A packet begins with a preamble, which consists of a variable number $ N_{\text{pr}} $ of upchirps, i.e., $ N_{\text{pr}} $ consecutive $ \mathbf{x}_{\text{ref}} $ symbols. After the preamble, the packet contains two network identifier symbols and $ 2 \nicefrac{1}{4} $ frequency synchronization symbols~\cite{Robyns2018}. The packet continues with an optional PHY header, which contains information about the length of the packet, the code rate, the presence of a cyclic redundancy check (CRC), and a checksum. The last part of the packet is the payload, which has a variable length with a maximum of 255 bytes, along with an optional 16-bit CRC of the payload bits~\cite{SX127x}. We focus on the FER for the payload part of the packet, which contains the actual transmitted data.

\section{LoRa Coded Frame Error Rate Under AWGN} \label{sec:FER}
Each LoRa payload consists of multiple interleaved Hamming codewords. Thus, in this section we first derive the codeword error rate (CWER) using existing results for the SER and then we use the CWER to derive the FER.

\subsection{Codeword Error Rate} \label{sec:CWER}

The uncoded symbol error probability $P_{s}$ is defined as $ P_{s} \triangleq P(\hat{s} \neq s) $. An approximation that can be used to efficiently evaluate the aforementioned probability under AWGN was derived in~\cite{Elshabrawy2018}. With our definition of the SNR, the symbol error probability can be calculated as
\begin{align}
  P_{s} & \approx Q\left(\frac{\sqrt{\text{SNR}} - \left((H_{N-1})^2 - \frac{\pi^2}{12}\right)^{1/4}}{\sqrt{H_{N-1} - \sqrt{(H_{N-1})^2 - \frac{\pi^2}{12}} + 0.5}}\right), \label{eq:SER_approximation}
\end{align}
where $Q(\cdot)$ denotes the Q-function and $H_n = \sum_{k=1}^n\frac{1}{k}$ denotes the $n$-th harmonic number. The term $(H_{N-1})^2$ in the above equation stems from the fact that, in the estimation of $\hat{s}$, $N-1$ symbols can be mistakenly chosen instead of $s$. When a symbol error happens due to AWGN only, on average half of the $\text{SF}$ symbol bits will be erroneous, independently of whether Gray mapping is used or not~\cite{Elshabrawy2018}. Thus, for the uncoded bit error probability $ P_{\text{b}} $ we have 
\begin{align}
   P_{b} & =  0.5\cdot P_{s} .
   \label{eq:BER}
\end{align}
The codeword error probability $ P_{\text{cw}} $ is defined as the probability that the Hamming decoder output decision $ \mathbf{\hat{c}} $ does not equal the Hamming encoder output codeword $\mathbf{c}$, i.e.,
\begin{align}
   P_{\text{cw}} & \triangleq  P(\hat{\mathbf{c}}\neq \mathbf{c}) .
\end{align}
The Hamming distance between two vectors $ \mathbf{w}_{1} $ and $ \mathbf{w}_{2} $, denoted by $ d(\mathbf{w}_{1},\mathbf{w}_{2}) $, is defined as the number of locations where $ \mathbf{w}_{1} $ and $ \mathbf{w}_{2} $ differ. Thus, if $ \mathbf{v} $ is the Hamming decoder input vector, then $ P_{\text{cw}} $ for the $(4,7)$ and $(4,8)$ Hamming codes (which can correct all one-bit errors) assuming that the decoder declares a failure when more than one error is detected can be equivalently defined as the probability of the event that $ \mathbf{v} $ has \emph{at least} two erroneous bits as
\begin{align}
   P_{\text{cw}} & = P(\{d(\mathbf{v},\mathbf{c}) \geqslant 2\}).
   \label{eq:CWER_78}
\end{align}

\begin{figure}
	\centering
	\includegraphics[width=0.50\textwidth]{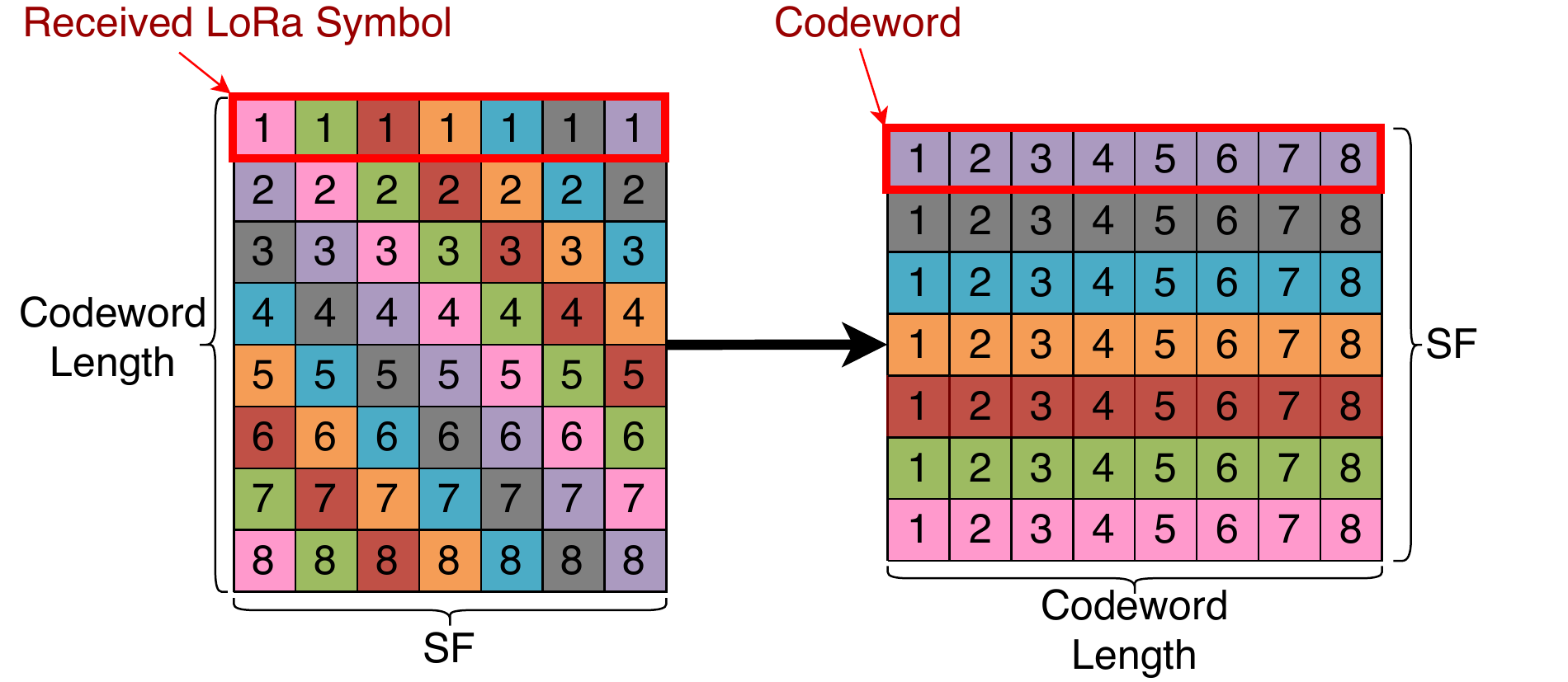}
	\caption{LoRa deinterleaving for $ \text{SF}=7  $ and a $ (4,8) $ Hamming code.}
	\label{fig:Interleaver}
	\vspace{-0.25cm}
\end{figure}

Fig.~\ref{fig:Interleaver} illustrates LoRa deinterleaving process. Each row of the matrix before deinterleaving corresponds to the $\text{SF}$ bits of a demodulated LoRa symbol  $ \mathbf{\hat{s}}_{i} $, $ i\in \{1, \hdots, n\} $, while each row of the matrix after deinterleaving corresponds to the $n$ bits of $ \mathbf{v}_{j} $, $j \in \{1,\hdots,\text{SF}\}$, at the Hamming decoder input.

When a symbol error happens, e.g., $ \mathbf{\hat{s}}_{1} \neq \mathbf{s}_{1} $, on average half of the bits of row $1$ of the matrix before deinterleaving will be erroneous. These bit errors are generally correlated because they come from the same symbol error, but the deinterleaver effectively removes this correlation by distributing each bit error of $\hat{\mathbf{s}}_{1}$ to distinct codewords, as seen in Fig.~\ref{fig:Interleaver}. If another symbol error occurs, e.g., $ \hat{\mathbf{s}}_{2} \neq \mathbf{s}_{2} $, the bit errors due to $ \hat{\mathbf{s}}_{2} $ are again generally correlated but they are independent from the bit errors of $ \hat{\mathbf{s}}_{1} $. This observation can be generalized to any number of symbol errors. As a result, the bit errors after the deinterleaver in every input codeword of the Hamming decoder $ \mathbf{v}_{j} $ are independent and identically distributed (iid) with probability of bit error $ P_{b} $ given by~\eqref{eq:BER}.
Thus, we can re-write the CWER in~\eqref{eq:CWER_78} as
\begin{align}
   P_{\text{cw}}
                 & = 1{-}\left( (1{-}P_{b})^{n} + \binom{n}{1}P_{b}(1{-}P_{b})^{n-1} \right).
                 \label{eq:CWER_final_78}
\end{align}

\subsection{Frame Error Rate} \label{sec:FER_AWGN}

We denote the number of payload symbols in a LoRa packet by $ N_{\text{pl}} $. By construction, $ N_{\text{pl}} $ is an integer multiple of the codeword length $n$. The number of codewords in the payload is $N_{\text{cw}} = \frac{N_{\text{pl}}\text{SF}}{n}$, where each codeword has a CWER $ P_{\text{cw}} $ given by~\eqref{eq:CWER_final_78}. Let $ \mathbf{C}_{\text{bl}} \in \{0,1\}^{\text{SF} \times n} $ be a matrix containing a block of $\text{SF}$ transmitted codewords $ \mathbf{c}_{i} $, $i \in \{1,\dots,\text{SF}\} $, that are input to one block of the interleaver. Let also $ \mathbf{V}_{\text{bl}} \in \{0,1\}^{\text{SF} \times n}$ be a matrix containing a block of $\text{SF}$ codewords that are the output of one deinterleaving block (and also the input to the Hamming decoder). Finally, let $ \hat{\mathbf{C}}_{\text{bl}} \in \{0,1\}^{\text{SF} \times n} $ be a matrix containing a block of $\text{SF}$ estimated codewords $ \hat{\mathbf{c}}_{i} $, $i \in \{1,\dots,\text{SF}\} $, after the Hamming decoder. The probability $ P\left(\hat{\mathbf{C}}_{\text{bl}} = \mathbf{C}_{\text{bl}}\right) $ corresponds to the probability that all the $ \text{SF} $ decoded codewords of one deinterleaver block are correctly decoded, which is given by
{\small
\begin{align}
	P\left(\hat{\mathbf{C}}_{\text{bl}}{=}\mathbf{C}_{\text{bl}}\right) & = \prod_{i=1}^{\text{SF}} P\left(\hat{\mathbf{c}_{i}} = \mathbf{c}_{i} | \mathbf{\hat{c}}_{1} = \mathbf{c}_{1},\hdots,\mathbf{\hat{c}}_{i-1} = \mathbf{c}_{i-1}\right).
\label{eq:joint}
\end{align}}%
Let $ P_{\text{cw}}^{(i)} $ denote the conditional codeword error probability given that all previous $i-1$ codewords in the block were decoded correctly, i.e.,
\begin{align}
P_{\text{cw}}^{(i)} = P(\hat{\mathbf{c}}_{i} \neq \mathbf{c}_{i} | \hat{\mathbf{c}}_{1} = \mathbf{c}_{1}, \dots ,\hat{\mathbf{c}}_{i-1} = \mathbf{c}_{i-1}). \label{eq:conditionalCWEP}
\end{align}

\begin{figure*}[t]
	\centering
	\begin{align}
		P\left(\hat{\mathbf{C}}_{\text{pl}}\neq \mathbf{C}_{\text{pl}}\right) & = 1 - \left(\prod_{i=1}^{\text{SF}}\left( 1-P_{b}^{(i)}\right)^{n} + \binom{n}{1}P_{b}^{(i)}\left(1-P_{b}^{(i)}\right)^{n-1} \right)^\frac{N_{\text{pl}}}{n}
\label{eq:payloadERapproxfull}
	\end{align}
	\hrule
\end{figure*}
Using~\eqref{eq:conditionalCWEP}, we can re-write~\eqref{eq:joint} as
\begin{align}
P\left(\hat{\mathbf{C}}_{\text{bl}} = \mathbf{C}_{\text{bl}}\right) = \prod_{i=1}^{\text{SF}}\left(1-P_{\text{cw}}^{(i)}\right).
\label{eq:joint_simpl}
\end{align}
Let $ \mathbf{C}_{\text{pl}} \in \{0,1\}^{N_{\text{cw}} \times n} $ be a matrix containing all the transmitted codewords in the payload and $ \hat{\mathbf{C}}_{\text{pl}} $ be a matrix containing all the decoded payload codewords. Since the payload contains $ \nicefrac{N_{\text{pl}}}{n} $ deinterleaving blocks which are independent from each other, using~\eqref{eq:joint_simpl} we can obtain the overall FER as
\begin{align}
P\left(\hat{\mathbf{C}}_{\text{pl}}\neq \mathbf{C}_{\text{pl}}\right) & = 1 - \left(\prod_{i=1}^{\text{SF}}\left(1-P_{\text{cw}}^{(i)}\right)\right)^\frac{N_{\text{pl}}}{n}.
\label{eq:payloadER}
\end{align}

As explained in Section~\ref{sec:CWER}, a correctly decoded codeword $ \hat{\mathbf{c}}_{i} $ may come from a decoder input $\mathbf{v}_{i}$ that has up to one bit error. This single non-catastrophic bit-error can be in any of the $ n $ positions of $ \mathbf{v}_{i} $. As a result, the computation of the product in~\eqref{eq:payloadER}, becomes a cumbersome task, since the number of all possible non-catastrophic bit-error patterns inside a deinterleaver block is a large combinatorial quantity. For this reason, in the remainder of this section we derive two low-complexity approximations for~\eqref{eq:payloadER}.

\textbf{Approximation 1:} 
A simple approximation of~\eqref{eq:payloadER} can be obtained by ignoring the conditioning in~\eqref{eq:conditionalCWEP} to obtain
\begin{align}
P\left(\hat{\mathbf{C}}_{\text{pl}}\neq \mathbf{C}_{\text{pl}}\right) \approx 1 - \left(1-P_{\text{cw}}\right)^\frac{N_{\text{pl}}\text{SF}}{n}.
\label{eq:joint_approx_1}
\end{align}
Since conditioning on the event that all previous $i-1$ codewords were decoded correctly intuitively decreases the probability that the $i$-th codeword will be erroneous, we expect that Approximation 1, which ignores this conditioning, overestimates the true FER.

\textbf{Approximation 2:} Let us now define a more elaborate approximation for~\eqref{eq:payloadER}, which has slightly higher computational complexity but gives more accurate results. To this end, in the conditional codeword error probability $P_{\text{cw}}^{(i)}$ of~\eqref{eq:conditionalCWEP}, we approximate the event that all previous $i-1$ codewords in the block were decoded correctly (i.e., $\{\hat{\mathbf{c}}_{1} = \mathbf{c}_{1}, \dots ,\hat{\mathbf{c}}_{i-1} = \mathbf{c}_{i-1}\}$) with the event that all previous $i-1$ codewords in the block did not contain any error at all (i.e., $\{\hat{\mathbf{v}}_{1} = \mathbf{c}_{1}, \dots ,\hat{\mathbf{v}}_{i-1} = \mathbf{c}_{i-1}\}$). Therefore, for all the previously decoded $i-1$ codewords in the block, we ignore the case where a single bit error has occurred. The conditional codeword error rate in~\eqref{eq:conditionalCWEP} can then be written as
\begin{align}
P_{\text{cw}}^{(i)} \approx P(\hat{\mathbf{c}}_{i} \neq \mathbf{c}_{i} | \mathbf{v}_{1} = \mathbf{c}_{1}, \dots ,\mathbf{v}_{i-1} = \mathbf{c}_{i-1}). \label{eq:conditionalCWEP_APP2}
\end{align}
Each conditional codeword error probability $ P_{\text{cw}}^{(i)} $ in~\eqref{eq:conditionalCWEP_APP2} can be interpreted as the codeword error probability of a decoder that has the additional side-information that $(i-1)$ out of the $n$ bit positions in $\mathbf{v}_{i}$ are guaranteed to be error-free. As such, we expect this approximation to result in slightly lower FER values compared to~\eqref{eq:payloadER}. However, we need to find an expression to calculate the conditional probability of~\eqref{eq:conditionalCWEP_APP2}.

Let $ P_{b}^{(i)} $ denote the conditional bit error probability $ P_{b}^{(i)} = 0.5 P_{s}^{(i)}$, where $ P_{s}^{(i)} $ is the conditional symbol error probability
\begin{align}
P_{s}^{(i)} = P(\hat{s} \neq s | \hat{b}_{1} = b_{1}, \dots ,\hat{b}_{i-1} = b_{i-1}),\label{eq:conditionalSEP_APP2}
\end{align}
where $ \hat{b}_{i} = b_{i} $ denotes that the $i$-th bit of symbol $s$ is estimated correctly. The additional information included in the conditional symbol error probability of~\eqref{eq:conditionalSEP_APP2}, compared to the unconditional one, is that a potential demodulation error can only be in $2^{\text{SF}-i-1} - 1$ DFT bins instead of the $N-1$ DFT bins with possible errors in~\eqref{eq:SER_approximation}. As such, the conditional symbol error probability $ P_{s}^{(i)} $ can be calculated by adapting the harmonic number in~\eqref{eq:SER_approximation} as follows
\begin{align}
P_{s}^{(i)} & \approx Q\left(\frac{\sqrt{\text{SNR}} {-} \left((H_{\frac{N}{2^{i-1}}-1})^2 {-} \frac{\pi^2}{12}\right)^{1/4}}{\sqrt{H_{\frac{N}{2^{i-1}}-1} {-} \sqrt{(H_{\frac{N}{2^{i-1}}-1})^2 {-} \frac{\pi^2}{12}} {+} 0.5}}\right). \label{eq:cond_SER_approximation}
\end{align}
Therefore, each $ P_{\text{cw}}^{(i)} $ in~\eqref{eq:payloadER} can be calculated as
\begin{align}
   P_{\text{cw}}^{(i)} & \approx 1 - \left( \left(1{-}P_{b}^{(i)}\right)^{n} + \binom{n}{1}P_{b}^{(i)}\left(1{-}P_{b}^{(i)}\right)^{n-1} \right). \label{eq:CWER_cond}
\end{align}
Finally, our second approximation for the FER can be obtained by replacing \eqref{eq:CWER_cond} in \eqref{eq:payloadER}, as shown in~\eqref{eq:payloadERapproxfull}.

\section{LoRa Coded Frame Error Rate Under AWGN and Carrier Frequency Offset} \label{sec:FER_AWGN_CFO}

In this section, we first derive an expression for the coded LoRa BER under AWGN and carrier frequency offset (CFO). We then use this initial result to explain how the coded FER can be calculated in the presence of both AWGN and CFO.

\subsection{Distribution of the Decision Metric}

In order to derive the coded FER under AWGN in Section~\ref{sec:FER}, we have used the symbol and bit error probability expressions from~\cite{Elshabrawy2018}. In order to derive the coded FER under CFO in this section, we first need an expression for the symbol and bit error probabilities of LoRa under residual CFO. Since such expressions do not exist in the literature, in the current subsection we introduce the necessary modeling that will allow us to find the uncoded symbol error probability as well as the coded bit error probability in Section~\ref{sec:SER_CFO}.

Let $ f_{c_1} $ and $ f_{c_2} $ be the carrier frequencies that are used during up- and down-conversion, respectively. The carrier frequency offset is the difference \mbox{$\Delta f_{c} = f_{c_1} - f_{c_2}$}. Thus, in the presence of CFO, the signal model becomes
\begin{align}
  y[n] & =  c[n]x[n] + z[n], \;\; n \in \mathcal{S}, \label{eq:lora_rx_cfo}
\end{align}
where $x[n]$ is the signal of interest, $ c[n] = e^{j2\pi n \frac{\Delta f_{c}}{f_{s}}}$ is the CFO term, and $z[n] \sim \mathcal{N}(0,\sigma^2)$ is AWGN. The demodulation of $y[n]$ at the receiver yields
\begin{align}
  \pmb{\mathcal{Y}}  & = \text{DFT}\left( \mathbf{c} \odot \mathbf{x} \odot \mathbf{x}_{\text{ref}}^{*}\right) + \text{DFT}\left(\mathbf{z} \odot \mathbf{x}_{\text{ref}}^{*}\right),
\end{align}
where $\mathbf{c} = \begin{bmatrix} c[0] & \hdots & c[N-1] \end{bmatrix}$.
We call $\text{DFT}\left(\mathbf{c} \odot \mathbf{x} \odot \mathbf{x}_{\text{ref}}^{*}\right)$ the \emph{CFO pattern}. The CFO pattern depends on the symbol value $s$ and on the CFO value $ \Delta f_{c} $. As explained in~\cite{Ghanaatian2019}, the CFO results in a frequency shift which can be modeled with an integer part $L$ and a fractional part $\lambda$, where $ L + \lambda = \frac{\Delta f_{c}N}{f_{s}}$. The fractional part $\lambda$ is the fraction of the CFO relative to the distance between two adjacent DFT bins. In this work, we consider \emph{residual CFO} after the estimation and correction procedures and we limit our study to the case where $L=0$ and $-0.5 \leq \lambda \leq 0.5$. If the residual CFO is so large that $L \neq 0$, then all symbols will anyway be erroneously demodulated.

\begin{figure*}
\small
\begin{align} 
P\left(\hat{s}\neq s|s,\lambda\right) & = P\left(\bigcup\limits_{i \in \mathcal{D}}\{|Y'_i| > |Y'_s|\}| s,\lambda\right) {+} P\left(\bigcup\limits_{j \in \mathcal{R}}\{|Y'_j| > |Y'_s|\}| s,\lambda\right) {-} P\left(\{\bigcup\limits_{i \in \mathcal{D}}\{|Y'_i| > |Y'_s|\} \cap \bigcup\limits_{j \in \mathcal{R}}\{|Y'_j| > |Y'_s|\}\}| s,\lambda\right) \label{eq:SER_three_terms}
\end{align}
\hrule
\end{figure*}

Let  $ \pmb{\mathcal{R}}_{k} $ denote the value of the CFO pattern at frequency bin $k$, i.e., $\pmb{\mathcal{R}}_{k} =  \text{DFT}\left(\mathbf{c} \odot \mathbf{x} \odot \mathbf{x}_{\text{ref}}^{*}\right)[k], \; k \in \mathcal{S}$. Using the definition of the Fourier transform and arguments that are similar to those in~\cite{Afisiadis2019b}, we have $ \pmb{\mathcal{R}}_{k} = A_{k} e^{-j\theta_{k}}$ where
\begin{align}
A_{k} & = \frac{\sin \left( \frac{\pi}{N} (s{-}k{+}\lambda)N \right)}{\sin \left( \frac{\pi}{N} (s{-}k{+}\lambda) \right)} \;\; \text{ and } \;\; \theta_{k} = \frac{\pi}{N} \left( s{-}k{+}\lambda \right)(N{-}1). \label{eq:A}
\end{align}
From \eqref{eq:A}, it can be seen that, in the absence of CFO (i.e., $\lambda=0$) we have $|\pmb{\mathcal{R}}_{k}|=0$, for any $k \in \mathcal{S}/s$, and $|\pmb{\mathcal{R}}_{s}|=N$. On the other hand, if $\lambda \neq 0$, a part of the energy of bin $s$ is spread across all $N$ bins.

We define $Y'_k = \frac{Y_k}{\sigma}$ for $k \in \mathcal{S}$, which can be used in~\eqref{eq:retrieved_symbol_dft} instead of $Y_k$. For a specific $ \lambda $ and symbol $s$, combining the CFO pattern with the AWGN leads to the demodulation metric $|Y'_k|$ used in~\eqref{eq:retrieved_symbol_dft} which is distributed according to~\cite{Afisiadis2019b}
\begin{align}
  |Y'_k| \;{\sim}\;
    f_{\text{Ri}}\left(y;\frac{|\pmb{\mathcal{R}}_{k}|}{\sigma},1\right), \;\; k \in \mathcal{S}.
    \label{eq:formula_Riceks_noise_cfo}
\end{align}

\subsection{Uncoded Symbol Error Rate} \label{sec:SER_CFO}
A symbol error occurs if and only if any of the $|Y'_k|$ values for $k \in \mathcal{S}/s$ exceeds the value of $|Y'_s|$, i.e.,
\begin{align} \label{eq:SER_general}
  \small
	P\left(\hat{s}\neq s|s,\lambda\right) & = P\left(\bigcup\limits_{k \in \mathcal{S}/s}\{|Y'_k| > |Y'_s|\}| s,\lambda\right).
\end{align}
As can be deduced from~\eqref{eq:A}, the DFT bins adjacent to bin $s$ have considerably higher energy due to the fractional offset $\lambda$ compared to the remaining bins, therefore, when CFO and AWGN are present the adjacent bins are generally more prone to causing symbol errors. However, in the case where the SNR is very low, all the bins in the set $\mathcal{S}/s$ still have comparable symbol error probabilities. In order to model the symbol error rate for a large range of SNRs, we rewrite the symbol error probability in~\eqref{eq:SER_general} in the equivalent form shown in \eqref{eq:SER_three_terms}, which separates the set $\mathcal{D} = \{s-1,s+1\}$ of the two adjacent bins, from the rest of the bins in the set $\mathcal{R} = \mathcal{S}/\{s-1,s,s+1\} $. The third term of~\eqref{eq:SER_three_terms} is typically small and can be ignored.

Due to the Gray mapping, a symbol error that mistakes $s$ for one of the symbols in $\mathcal{D}$ only causes one bit error. On the other hand, when a symbol error happens that mistakes $s$ for one of the symbols in $\mathcal{R}$, on average half of the bits are wrong. Therefore, the bit error probability can be written as

\begin{align} \label{eq:BER_twoterms}
\small
P\left(\hat{b}\neq b|s,\lambda\right) & \approx \frac{1}{\text{SF}} P\left(\{|Y'_{\max,\mathcal{D}}| > |Y'_s|\}| s,\lambda\right) \nonumber\\
&+ \frac{1}{2}P\left(\{|Y'_{\max,\mathcal{R}}| > |Y'_s|\}| s,\lambda\right),
\end{align}
where $|Y'_{\max,\mathcal{D}}| = \max _{i \in \mathcal{D}}|Y'_i|$ and $|Y'_{\max,\mathcal{R}}| = \max _{j \in \mathcal{R}}|Y'_j|$. The first probability term in~\eqref{eq:BER_twoterms} is given by
{\small
\begin{align} \label{eq:first_term}
P\left(\{|Y'_{\max,\mathcal{D}}| {>} |Y'_s|\}| s,\lambda\right) & = 1{-}\int_{y=0}^{+\infty} f_{\text{Ri}}\left(y;v_{s},1\right) F_{|Y'_{\max,\mathcal{D}}|} (y) dy,
\end{align}
}%
where $v_{s} = \nicefrac{|R_s|}{\sigma}$ is the location parameter for bin $s$. The second term in~\eqref{eq:BER_twoterms} can be obtained by replacing $\mathcal{D}$ with $\mathcal{R}$. The CDF of the n-th order statistic (i.e., the CDF of the maximum) is known to be $F_{n}(x) = P(X_1 < x) P(X_2 < x) \dots P(X_n < x)$. Due to the conditioning on $\lambda$, each $ |Y'_m| $ for $m\in \mathcal{R} $ is independent from any other $ |Y'_n| $ for $n \in \mathcal{D} / m $. Thus, the CDF of the maximum bin for the set $\mathcal{D}$ is
\begin{equation}
	F_{|Y'_{\max,\mathcal{D}}|} (y) = \prod_{j \in \mathcal{D}} F_{\text{Ri}}(y;v_j,1), \label{eq:cdf_back_bins}
\end{equation}
where $v_j = \nicefrac{|R_j|}{\sigma}$. The CDF $F_{|Y'_{\max,\mathcal{R}}|}$ can be obtained accordingly. Finally, $s$ is uniformly distributed in  $\mathcal{S}$, so that
\begin{align}
P_{b|\lambda} = P\left(\hat{b}\neq b|\lambda\right) & = \frac{1}{N}\sum_{s=0}^{N-1} P\left(\hat{b}\neq b|s,\lambda\right). \label{eq:ser_cfo_full}
\end{align}

\subsection{Complexity Reduction}

We can see from~\eqref{eq:A} that all CFO patterns for all values of $s \in \mathcal{S}$ contain the same set of frequency bin magnitudes $|\pmb{\mathcal{R}}_{k}|$, $ k\in \mathcal{S} $, but are circularly shifted. This circular shift does not change the distribution of $ |Y'_{\max}|$, thus the probability of $|Y'_{\max,\mathcal{D}}| > |Y'_s|$ as well as the probability of $|Y'_{\max,\mathcal{R}}| > |Y'_s|$ are not affected. Moreover, the errors in the adjacent bins always give one-bit errors due to the Gray mapping, independently of the circular shift. Therefore, the CFO patterns for all $s \in \mathcal{S}$ result in exactly the same bit error probability $P(\hat{b}\neq b|s,\lambda)$. This means that it is sufficient to compute the CFO pattern for any single symbol $s$ for the evaluation of the bit error probability $P_{b|\lambda}$ given in~\eqref{eq:ser_cfo_full}, thus reducing the complexity of evaluating~\eqref{eq:ser_cfo_full} by a factor of $N$.

\subsection{Frame Error Rate under AWGN and CFO}
We follow a similar approach to the AWGN case in order to approximate the coded FER under CFO. First, we approximate the CWER under CFO as
\begin{align}
P_{\text{cw,cfo}}
& = 1 - \left( (1{-}P_{b|\lambda})^{n} + \binom{n}{1}P_{b|\lambda}(1{-}P_{b|\lambda})^{n-1} \right).
\label{eq:CWER_CFO}
\end{align}
Following the same reasoning as Approximation 1 of Section~\ref{sec:FER_AWGN}, the FER under CFO can then be written as
\begin{align}
P\left(\hat{\mathbf{C}}_{\text{pl,cfo}}\neq \mathbf{C}_{\text{pl,cfo}}\right) \approx 1 - \left(1-P_{\text{cw,cfo}}\right)^\frac{N_{\text{pl}}\text{SF}}{n}.
\label{eq:fer_cfo}
\end{align}

\section{Numerical Results} \label{sec:results}

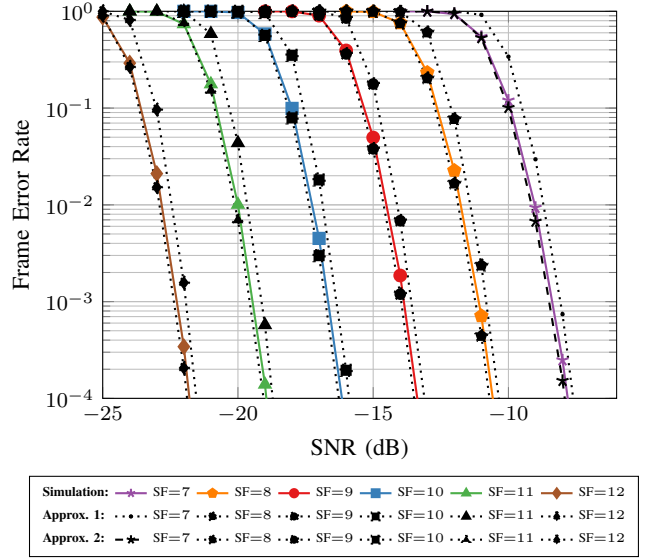
\begin{figure}
  \centering
  \begin{tikzpicture}

	\small

	\begin{semilogyaxis}[
		width = \figurewidth\columnwidth,
		height = \figureheight\columnwidth,
		xlabel = {SNR (dB)},
		ylabel = {Frame Error Rate},
		label style={font=\small},
    tick label style={font=\footnotesize},
		ylabel near ticks,
		xlabel near ticks,
		xmin = -25, xmax = -6,
		ymin = 1e-4, ymax = 1,
		grid = both,
		legend image post style={scale=0.7},
		legend style={at={(0.45,-0.2)},anchor=north,font=\tiny},
		legend cell align={left},
		legend columns={7},
	]

		\addlegendimage{empty legend}
		\addlegendentry{\textbf{Simulation:}}

		\addplot[Set1-7-4, thick, solid, mark=star, mark options={scale=1.2}] table[x index=0, y index = 1] {figs/data/PER_RES_SF7_payloadLen32_preambleLen8_preambleThresh3_PPM8_Interl_Or_Hamming.dat};
		\addlegendentry{SF${=}7$}
		\addplot[Set1-7-5, thick, solid, mark=pentagon*, mark options={scale=1.2}] table[x index=0, y index = 1] {figs/data/PER_RES_SF8_payloadLen32_preambleLen8_preambleThresh3_PPM8_Interl_Or_Hamming.dat};
		\addlegendentry{SF${=}8$}
		\addplot[Set1-7-1, thick, solid, mark=*, mark options={scale=1.1}] table[x index=0, y index = 1] {figs/data/PER_RES_SF9_payloadLen32_preambleLen8_preambleThresh3_PPM8_Interl_Or_Hamming.dat};
		\addlegendentry{SF${=}9$}
		\addplot[Set1-7-2, thick, solid, mark=square*, mark options={scale=1.05}] table[x index=0, y index = 1] {figs/data/PER_RES_SF10_payloadLen32_preambleLen8_preambleThresh3_PPM8_Interl_Or_Hamming.dat};
		\addlegendentry{SF${=}10$}
		\addplot[Set1-7-3, thick, solid, mark=triangle*, mark options={scale=1.2}] table[x index=0, y index = 1] {figs/data/PER_RES_SF11_payLen32_prLen2_prMargin3_CWLen8_0.dat};
		\addlegendentry{SF${=}11$}
		\addplot[Set1-7-6, thick, solid, mark=diamond*, mark options={scale=1.3}] table[x index=0, y index = 1] {figs/data/PER_RES_SF12_payLen32_prLen2_prMargin3_CWLen8_0.dat};
		\addlegendentry{SF${=}12$}

		\addlegendimage{empty legend}
		\addlegendentry{\textbf{Approx. 1:}}
		\addplot[black, thick, dotted, mark=star, mark options={scale=1.2}] table[x index=0, y index = 1] {figs/data/APP_simpl_PER_AWGN_SF7.dat};
		\addlegendentry{SF${=}7$}
		\addplot[black, thick, dotted, mark=pentagon*, mark options={scale=1.2}] table[x index=0, y index = 1] {figs/data/APP_simpl_PER_AWGN_SF8.dat};
		\addlegendentry{SF${=}8$}
		\addplot[black, thick, dotted, mark=*, mark options={scale=1.1}] table[x index=0, y index = 1] {figs/data/APP_simpl_PER_AWGN_SF9.dat};
		\addlegendentry{SF${=}9$}
		\addplot[black, thick, dotted, mark=square*, mark options={scale=1.05}] table[x index=0, y index = 1] {figs/data/APP_simpl_PER_AWGN_SF10.dat};
		\addlegendentry{SF${=}10$}
		\addplot[black, thick, dotted, mark=triangle*, mark options={scale=1.2,solid}] table[x index=0, y index = 1] {figs/data/APP_simpl_PER_AWGN_SF11.dat};
		\addlegendentry{SF${=}11$}
		\addplot[black, thick, dotted, mark=diamond*, mark options={scale=1.3}] table[x index=0, y index = 1] {figs/data/APP_simpl_PER_AWGN_SF12.dat};
		\addlegendentry{SF${=}12$};

		\addlegendimage{empty legend}
		\addlegendentry{\textbf{Approx. 2:}}
		\addplot[black, thick, dashed, mark=star, mark options={scale=1.2}] table[x index=0, y index = 1] {figs/data/APP_PER_AWGN_SF7.dat};
		\addlegendentry{SF${=}7$}
		\addplot[black, thick, dotted, mark=pentagon*, mark options={scale=1.2}] table[x index=0, y index = 1] {figs/data/APP_PER_AWGN_SF8.dat};
		\addlegendentry{SF${=}8$}
		\addplot[black, thick, dotted, mark=*, mark options={scale=1.1}] table[x index=0, y index = 1] {figs/data/APP_PER_AWGN_SF9.dat};
		\addlegendentry{SF${=}9$}
		\addplot[black, thick, dotted, mark=square*, mark options={scale=1.05}] table[x index=0, y index = 1] {figs/data/APP_PER_AWGN_SF10.dat};
		\addlegendentry{SF${=}10$}
		\addplot[black, thick, dotted, mark=triangle*, mark options={scale=1.2}] table[x index=0, y index = 1] {figs/data/APP_PER_AWGN_SF11.dat};
		\addlegendentry{SF${=}11$}
		\addplot[black, thick, dotted, mark=diamond*, mark options={scale=1.3}] table[x index=0, y index = 1] {figs/data/APP_PER_AWGN_SF12.dat};
		\addlegendentry{SF${=}12$};

	\end{semilogyaxis}

\end{tikzpicture}%
  \caption{Frame error rate of the LoRa modulation under AWGN for all supported spreading factors $\text{SF} \in \left\{7,\hdots,12\right\}$.}
  \label{fig:perawgn}
  \vspace{-0.25cm}
\end{figure}

In Fig.~\ref{fig:perawgn}, we first compare the results of a Monte Carlo simulation for the LoRa FER for all possible $\text{SF} \in \left\{7,\hdots,12\right\}$ with the FER results obtained by using Approximation 1 and Approximation 2 under AWGN.
The payload length is chosen to be $ N_{\text{pl}} = 32 $ LoRa symbols. We observe that both approximations are quite accurate, especially at low FER values, but Approximation 2 is visibly better than Approximation 1. However, Approximation 1 only requires a single Q-function evaluation per SNR point, while Approximation 2 requires $\text{SF}$ Q-function evaluations per SNR point. Moreover, we observe that Approximation 1 slightly overestimates and Approximation 2 slightly underestimates the error rate, as expected from the discussion in Section~\ref{sec:FER}.

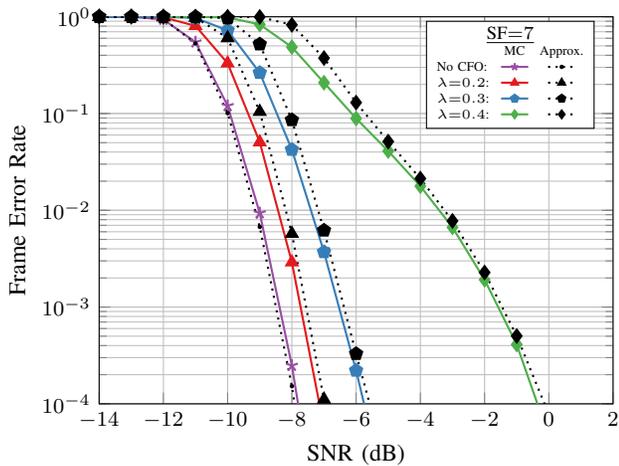
\begin{figure}
  \centering
  \begin{tikzpicture}

	\small

	\begin{semilogyaxis}[
		width = \figurewidth\columnwidth,
		height = \figureheight\columnwidth,
		xlabel = {SNR (dB)},
		ylabel = {Frame Error Rate},
		label style={font=\small},
    tick label style={font=\footnotesize},
		ylabel near ticks,
		xlabel near ticks,
		xmin = -14, xmax = 2,
		ymin = 1e-4, ymax = 1,
		grid = both,
		legend image post style={scale=0.6},
	]


		\addplot[Set1-7-4, thick, solid, mark=star, mark options={scale=1.2}] table[x index=0, y index = 1] {figs/data/PER_RES_SF7_payloadLen32_preambleLen8_preambleThresh3_PPM8_Interl_Or_Hamming.dat};
		\label{lambda0.0MC}
		\addplot[Set1-7-1, thick, solid, mark=triangle*, mark options={scale=1.2}] table[x index=0, y index = 1] {figs/data/RES_SF7_F32_tau0.20_CWLen8_Interl1_Gray1_0.dat};
		\label{lambda0.2MC}
		\addplot[Set1-7-2, thick, solid, mark=pentagon*, mark options={scale=1.2}] table[x index=0, y index = 1] {figs/data/RES_SF7_F32_tau0.30_CWLen8_Interl1_Gray1_0.dat};
		\label{lambda0.3MC}
		\addplot[Set1-7-3, thick, solid, mark=diamond*, mark options={scale=1.2}] table[x index=0, y index = 1] {figs/data/RES_SF7_F32_tau0.40_CWLen8_Interl1_Gray1_0.dat};
		\label{lambda0.4MC}

		\addplot[black, thick, dotted, mark=star, mark options={scale=1.2}] table[x index=0, y index = 1] {figs/data/APP_PER_AWGN_SF7.dat};
		\label{lambda0.0APP}
		\addplot[black, thick, dotted, mark=triangle*, mark options={scale=1.2, solid}] table[x index=0, y index = 1] {figs/data/APP_SF7_F32_lambda0.20_Gray1.dat};
		\label{lambda0.2APP}
		\addplot[black, thick, dotted, mark=pentagon*, mark options={scale=1.2, solid}] table[x index=0, y index = 1] {figs/data/APP_SF7_F32_lambda0.30_Gray1.dat};
		\label{lambda0.3APP}
		\addplot[black, thick, dotted, mark=diamond*, mark options={scale=1.2, solid}] table[x index=0, y index = 1] {figs/data/APP_SF7_F32_lambda0.40_Gray1.dat};
		\label{lambda0.4APP}

		\node [draw,fill=white,inner sep=2pt] at (rel axis cs: 0.8,0.85) {
		\tiny
		\begin{tabular}{lcc}
			\multicolumn{3}{c}{\scriptsize\underline{SF${=}7$}} \\
								& MC 								& Approx. \\
			No CFO:	 	& \ref{lambda0.0MC} 	&	\ref{lambda0.0APP} \\
			$\lambda {=} 0.2$:	 	& \ref{lambda0.2MC} 	&	\ref{lambda0.2APP} \\
			$\lambda {=} 0.3$: 	& \ref{lambda0.3MC} & \ref{lambda0.3APP} \\
			$\lambda {=} 0.4$: 	& \ref{lambda0.4MC} & \ref{lambda0.4APP}
		\end{tabular}};


	\end{semilogyaxis}

\end{tikzpicture}%
  \caption{LoRa FER under three different values of CFO for $\text{SF} = 7$.}
  \label{fig:perCFOsf7}
  \vspace{-0.25cm}
\end{figure}

In Fig.~\ref{fig:perCFOsf7}, we show the FER for a coded LoRa system with $\text{SF} = 7$ for three different values of the fractional CFO ($\lambda \in \{ 0.2$, $0.3$, $0.4$\}), as well as the case without any CFO. The payload length is chosen to be $ N_{\text{pl}} = 32 $ LoRa symbols. We observe that the impact of CFO is significant, especially for large $\lambda$ and at low error rates. Moreover, we can see that our derived approximation is quite accurate for all values of $\lambda$.

\begin{figure}
	\centering
	\begin{tikzpicture}

	\small

	\begin{semilogyaxis}[
		width = \figurewidth\columnwidth,
		height = \figureheight\columnwidth,
		xlabel = {SNR (dB)},
		ylabel = {Frame Error Rate},
		label style={font=\small},
    tick label style={font=\footnotesize},
		ylabel near ticks,
		xlabel near ticks,
		xmin = -25, xmax = -6,
		ymin = 1e-4, ymax = 1,
		grid = both,
		legend image post style={scale=0.7},
		legend style={at={(0.45,-0.2)},anchor=north,font=\tiny},
		legend cell align={left},
		legend columns={7},
	]

		\addlegendimage{empty legend}
		\addlegendentry{\textbf{Simulation:}}
		
		\addplot[Set1-7-4, thick, solid, mark=star, mark options={scale=1.2}] table[x index=0, y index = 1] {figs/data/RES_SF7_F32_tau0.20_CWLen8_Interl1_Gray1_0.dat};
		\addlegendentry{SF${=}7$}
		\addplot[Set1-7-5, thick, solid, mark=pentagon*, mark options={scale=1.2}] table[x index=0, y index = 1] {figs/data/PER_RES_SF8_F32_lambda0.20_CWLen8_Interl1_Gray1_Reverse1_Reduced0_0.dat};
		\addlegendentry{SF${=}8$}
		\addplot[Set1-7-1, thick, solid, mark=*, mark options={scale=1.1}] table[x index=0, y index = 1] {figs/data/PER_RES_SF9_F32_lambda0.20_CWLen8_Interl1_Gray1_Reverse1_Reduced0_0.dat};
		\addlegendentry{SF${=}9$}
		\addplot[Set1-7-2, thick, solid, mark=square*, mark options={scale=1.05}] table[x index=0, y index = 1] {figs/data/PER_RES_SF10_F32_lambda0.20_CWLen8_Interl1_Gray1_Reverse1_Reduced0_0.dat};
		\addlegendentry{SF${=}10$}
		\addplot[Set1-7-3, thick, solid, mark=triangle*, mark options={scale=1.2}] table[x index=0, y index = 1] {figs/data/PER_RES_SF11_F32_lambda0.20_CWLen8_Interl1_Gray1_Reverse1_Reduced0_0.dat};
		\addlegendentry{SF${=}11$}
		\addplot[Set1-7-6, thick, solid, mark=diamond*, mark options={scale=1.3}] table[x index=0, y index = 1] {figs/data/PER_RES_SF12_F32_lambda0.20_CWLen8_Interl1_Gray1_Reverse1_Reduced0_0.dat};
		\addlegendentry{SF${=}12$}
		
		\addlegendimage{empty legend}
		\addlegendentry{\textbf{Approx.:}}
		\addplot[black, thick, dotted, mark=star, mark options={scale=1.2}] table[x index=0, y index = 1] {figs/data/APP_SF7_F32_lambda0.20_Gray1.dat};
		\addlegendentry{SF${=}7$}
		\addplot[black, thick, dotted, mark=pentagon*, mark options={scale=1.2}] table[x index=0, y index = 1] {figs/data/APP_SF8_F32_lambda0.20_Gray1.dat};
		\addlegendentry{SF${=}8$}
		\addplot[black, thick, dotted, mark=*, mark options={scale=1.1}] table[x index=0, y index = 1] {figs/data/APP_SF9_F32_lambda0.20_Gray1.dat};
		\addlegendentry{SF${=}9$}
		\addplot[black, thick, dotted, mark=square*, mark options={scale=1.05}] table[x index=0, y index = 1] {figs/data/APP_SF10_F32_lambda0.20_Gray1.dat};
		\addlegendentry{SF${=}10$}
		\addplot[black, thick, dotted, mark=triangle*, mark options={scale=1.2,solid}] table[x index=0, y index = 1] {figs/data/APP_SF11_F32_lambda0.20_Gray1.dat};
		\addlegendentry{SF${=}11$}
		\addplot[black, thick, dotted, mark=diamond*, mark options={scale=1.3}] table[x index=0, y index = 1] {figs/data/APP_SF12_F32_lambda0.20_Gray1.dat};
		\addlegendentry{SF${=}12$};


		\addplot[lightgray, ultra thick, solid] table[x index=0, y index = 1] {figs/data/PER_RES_SF7_payloadLen32_preambleLen8_preambleThresh3_PPM8_Interl_Or_Hamming.dat};
\addplot[lightgray, ultra thick, solid] table[x index=0, y index = 1] {figs/data/PER_RES_SF8_payloadLen32_preambleLen8_preambleThresh3_PPM8_Interl_Or_Hamming.dat};
\addplot[lightgray, ultra thick, solid] table[x index=0, y index = 1] {figs/data/PER_RES_SF9_payloadLen32_preambleLen8_preambleThresh3_PPM8_Interl_Or_Hamming.dat};
\addplot[lightgray, ultra thick, solid] table[x index=0, y index = 1] {figs/data/PER_RES_SF10_payloadLen32_preambleLen8_preambleThresh3_PPM8_Interl_Or_Hamming.dat};
\addplot[lightgray, ultra thick, solid] table[x index=0, y index = 1] {figs/data/PER_RES_SF11_payLen32_prLen2_prMargin3_CWLen8_0.dat};
\addplot[lightgray, ultra thick, solid] table[x index=0, y index = 1] {figs/data/PER_RES_SF12_payLen32_prLen2_prMargin3_CWLen8_0.dat};

	\end{semilogyaxis}

\end{tikzpicture}%
	\caption{Frame error rate of the LoRa modulation under AWGN and CFO ($\lambda=0.2$) for all supported spreading factors $\text{SF} \in \left\{7,\hdots,12\right\}$. The thick gray lines show the FER when only AWGN is present for comparison.}
	\label{fig:perCFOallSF}
	\vspace{-0.25cm}
\end{figure}
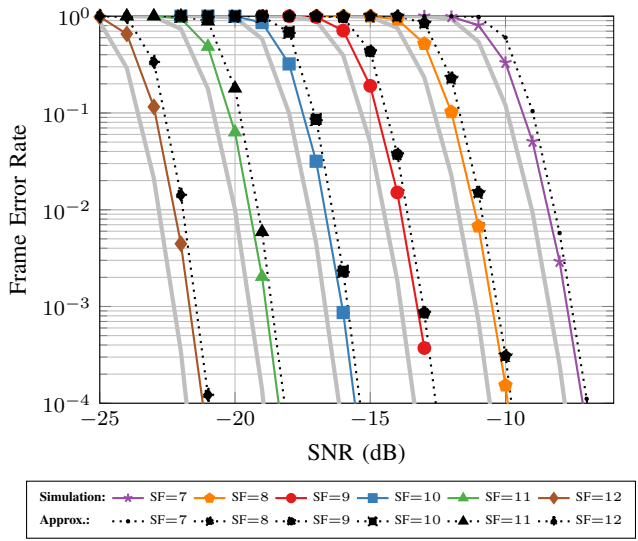

In Fig.~\ref{fig:perCFOallSF}, we compare the results of a Monte Carlo simulation for the LoRa FER for all possible $\text{SF} \in \left\{7,\hdots,12\right\}$ with the FER results obtained by using our derived approximation for $\lambda=0.2$. The FER under only AWGN is also included in the figure
with thick gray lines (taken from Fig.~\ref{fig:perawgn}). We note that the choice of a common fractional offset $\lambda$ for all spreading factors corresponds to different $\Delta f_{c}$ values for each SF. We observe that, for all spreading factors, a common fractional offset value leads to similar performance degradation and that our derived approximation is accurate.

\section{Conclusion} \label{sec:conclusion}
In this work, we analyzed and quantified the coded frame error rate performance of LoRa under AWGN and carrier frequency offset. Specifically, we first derived two low-complexity approximations for the coded LoRa FER under AWGN. The FER obtained by using the second approximation is shown to be within 0.2~dB of the FER obtained through a Monte Carlo simulation for all LoRa spreading factors and FERs down to $10^{-5}$. Moreover, we derived the FER under the impact of CFO and we showed that, thanks to the combination of Gray mapping, interleaving, and coding, LoRa is relatively robust to small values of residual CFO. Finally, we derived a low-complexity approximation for the LoRa FER under CFO which is no more than 0.5~dB away from the corresponding Monte Carlo simulation.

\section*{Acknowledgment}
The authors would like to thank Joachim Tapparel for useful discussions on implementation details of the LoRa PHY.

\bibliographystyle{IEEEtran}
\bibliography{IEEEabrv,refs}

\end{document}